\documentclass[a4paper,11pt]{article}

\usepackage{amsmath}
\usepackage{amsfonts}
\usepackage{amssymb}
\usepackage{lscape}
\usepackage{rotating}

\begin{document}

\title{Gravitational Radiation Theory and Light Propagation\footnote{to appear in: C. L\"ammerzahl, C.W.F. Everitt, F.W. Hehl (eds.): {\it Gyros, Clocks, and Interferometers: Testing Relativistic Gravity in Space}, Springer--Verlag, Berlin 2000}}

\author{Luc Blanchet$^{1}$, Sergei Kopeikin$^{2}$, and 
Gerhard Sch\"afer$^{3}$ \\ 
$^{1}$ D\'epartement d'Astrophysique Relativiste et de Cosmologie 
(CNRS), \\ 
Observatoire de Paris, 92195 Meudon Cedex, France
\\ 
$^{2}$ Department of Physics \& Astronomy,
University of Missouri-Columbia, \\ 
Physics Building 223,
Columbia, MO 65211, USA
\\ 
$^{3}$ Theoretisch--Physikalisches Institut,
Friedrich--Schiller--Universit\"at, \\
Max--Wien--Platz 1, 07743 Jena, Germany}

\maketitle             

\begin{abstract}
The paper gives an introduction to the gravitational radiation theory
of isolated sources and to the propagation properties of light rays 
in radiative gravitational fields.
It presents a theoretical study of the generation, 
propagation, back--reaction, and detection of gravitational waves 
from astrophysical sources. After reviewing the various 
quadrupole--moment laws for gravitational radiation in the 
Newtonian approximation, we show how to incorporate post--Newtonian 
corrections into the source multipole moments, the radiative multipole 
moments at infinity, and the back-reaction potentials. We further treat 
the light propagation in the linearized gravitational field outside 
a gravitational wave emitting source. The effects of 
time delay, bending of light, and moving source frequency shift 
are presented in terms of the gravitational lens potential. 
Time delay results are applied in the description of the procedure of 
the detection of gravitational waves. 

Pacs Numbers : 04.25.-g, 04.25.Nx
\end{abstract}

\section{Introduction}\index{gravitational waves}

It was only in the late fifties of the twentieth century that by the work of 
Hermann Bondi and Joseph Weber gravitational radiation entered 
the domain of physics. Before that time gravitational radiation was not 
considered to be of observational relevance and the gravitational radiation 
theory was not developed very deeply.

The supposed detection of gravitational radiation by J. Weber in the late 
sixties triggered strong and still on-going efforts both in the building of 
gravitational wave detectors and in the elaboration of the gravitational 
radiation theory, including investigations of the most reliable sources of 
detectable gravitational waves, calculations of wave forms, and analysis of 
data from detectors (cf. \cite{B91}). It turned out that coalescing neutron 
stars and/or stellar-mass black holes, together with gravitationally collapsing 
objects (type II supernovae), are the most relevant sources for detectable gravitational 
waves on Earth because they are strong and fit well to the frequency band of the 
Earth-based detectors which ranges from 10 Hz to 10 kHz. The strength of these 
sources is such high that several detection events per year might be expected   
in future fully developed detectors. The most sensitive 
Earth-based detectors will go into operation in the first few years 
of the new millenium. 
These are the laser-interferometric detectors in Germany, \index{GEO 600} GEO600, built by a 
German/British consortium, in Italy, VIRGO, built by a Italian/French 
consortium, the two LIGO detectors in the United States, and the TAMA300 
detector in Japan. There are several bar detectors already operating on 
Earth (ALLEGRO in the United States, AURIGA and NAUTILUS in Italy, EXPLORER 
at CERN, NIOBE in Australia). These detectors are being permanently upgraded 
and will be supplementing the measurements of the interferometric detectors
later. 
For the measurement of gravitational waves in the frequency range between
0.1 Hz and 0.1 mHz the space-borne laser-interferometric detector LISA is 
devised which is expected to be flown around 2010 by NASA/ESA. The 
astrophysical sources of the gravitational waves to be detected by this 
detector are a variety of orbiting stars (interacting
white dwarf binaries, compact binaries), orbiting massive black holes, 
as well as the formation and coalescence of supermassive black holes.
Stochastic gravitational waves from the early universe are expected to
exist in the whole measurable frequency range from $10^4$ Hz down 
to $10^{-18}$ Hz. The tools to possibly measure the primordial waves are 
the mentioned Earth-based and space-borne detectors, Doppler tracking,
pulsar timing, very long baseline interferometry, as well as the cosmic 
microwave background. 

On the theoretical side there are essentially two approaches which permit 
to investigate the properties of and to make predictions about 
gravitational waves from various sources. The first approach, that we can 
qualify as ``exact'', stays within the exact theory, solving or establishing 
theorems about the complete non-linear Einstein field equations. Within this 
approach one can distinguish the work dealing with exact solutions of the 
field equations in the form of plane gravitational waves, and especially 
colliding plane waves. Since the waves are never planar in nature, this work 
is not very relevant to real astrophysics, but its academic interest is 
important in that it permits notably the study of the appearance of 
singularities triggered by collisions of waves. Also within the exact 
approach, but more important for applications in astrophysics, is all the work concerned with the study of the asymptotic structure of the 
gravitational field of isolated radiating systems. The work on asymptotics 
started with the papers of Bondi {\it et al.} \cite{BBM62} and 
Penrose \cite{P65}. The second approach is much more general, in the sense 
that it is not restricted to any particular symmetry of the system, nor it is 
applicable only in the far region of the system. However, the drawback of this 
approach is that it is only {\it approximate} and essentially looks for the 
solutions of the Einstein field equations in the form of formal expansions 
when $c\to\infty$ (post-Newtonian approximation). This approximate 
post-Newtonian method can be applied to the study of all theoretical aspects of 
gravitational radiation: the equations of motion of the source including 
the gravitational radiation reaction (works of Einstein, Infeld, and 
Hoffmann \cite{EIH}, Chandrasekhar and Esposito \cite{CE70}, Burke and 
Thorne \cite{Bu71,BuTh70,MTW}, Ehlers \cite{Ehl80}, Papapetrou and 
Linet \cite{PapaL81}, Damour and Deruelle \cite{DD81a,D83a}, 
Sch\"afer \cite{S85}, Kopejkin \cite{Kop85}); the structure of the 
radiation field (work of Bonnor \cite{Bo59}, Thorne \cite{Th80}, Blanchet 
and Damour \cite{BD86}), and, more recently, accurate post-Newtonian wave 
generation formalisms \cite{BD89,DI91a,BD92,B95,WWi96}.

To the lowest, Newtonian order, the wave-generation formalism is called the 
quadrupole formalism, because as a consequence of the equality of the inertial 
and gravitational mass of all bodies the dominant radiating moment of any 
system is the (mass-type) quadrupole, which simply is at this approximation 
the standard Newtonian quadrupole moment. We are very much confident in using 
the theoretical framework of the post-Newtonian approximation 
because, marvellously enough, the framework of the Newtonian, 
quadrupole formalism has been checked by astronomical observations.

In fact, there are two observational tests of the validity of the quadrupole formalism. 
The first 
test concerns the famous Hulse--Taylor binary pulsar whose decrease of the orbital period $P_b$ by 
gravitational radiation is predicted from the quadrupole formula to be
\cite{PeM63,EH75,Wag75,D83b}
\begin{equation}
\label{1}
{\dot P}_b=-{192\pi\over 5c^5}\left({2\pi G\over P_b}\right)^{5/3}
{M_pM_c\over (M_p+M_c)^{1/3}}{1+{73\over 24}e^2+{37\over 96}
e^4\over (1-e^2)^{7/2}}\;,
\end{equation}
where $M_p$ and $M_c$ are the pulsar and companion masses, $e$ is the orbit eccentricity, 
and $G,c$ are the universal gravitational constant and the speed of light. Numerically, 
one finds ${\dot P}_b=-2.4 \times 10^{-12}$ sec/sec, in excellent agreement 
($0.35\%$ precision) with the 
observations by Taylor {\it et al.} \cite{TFMc79,T93}. 
The second test concerns the so-called cataclysmic variables. There we have  
binary systems in which a star 
filling its Roche lobe (the ``secondary'' with mass $M_2$) transfers mass onto a 
more massive white dwarf (the ``primary'' with mass $M_1>M_2$). From the 
formula for the angular momentum in Newtonian theory $J=G M_1 M_2 (a/GM)^{1/2}$
(where $M=M_1+M_2$), we deduce the secular evolution of the orbital semi-major 
radius $a$ (whatever may be the mechanism for the variation of $J$),
\begin{equation}
\label{2}
{{\dot a}\over a}={2{\dot J}\over J}-{2{\dot M_2}\over M_2}
\left(1-{M_2\over M_1}\right)\;,
\end{equation}
where ${\dot M}_2$ is the rate at which the secondary transfers mass to the 
primary (${\dot M}_2<0$). 
Since $M_1>M_2$, the mass transfer tends to increase the radius $a$ 
of the orbit, 
hence to increase the radius of the secondary's Roche lobe, and, thus, to 
stop the mass transfer. Therefore a long lived mass transfer is possible 
only if the system looses angular momentum to compensate for the 
increase of $a$. For cataclismic binaries with periods longer than about 
two hours, the loss of angular momentum is explained by standard astrophysical 
theory (interaction between the magnetic field and the stellar wind of the 
secondary). But for short-period binaries, with period less than about two hours, the 
only way to explain the loss of angular momentum is to invoke gravitational radiation. 
Now, from the quadrupole formula, we have
\begin{equation}
\label{3}
\left({{\dot J}\over J}\right)_{\rm GW}=-{32G^3\over 5c^5}{M_1M_2M\over 
a^4}\;.
\end{equation}
Inserting this into (\ref{2}) one can then predict what should be 
${\dot M}_2$ in order that $\dot{a}/a \sim 0$, and the result is in good 
agreement with the mass transfer measured from the X-rays observations of 
cataclysmic binaries.

Another important aspect of the theory of gravitational radiation, 
with obvious implications in astronomy, is the interaction of the 
gravitational wave field with photons. In this article we present the results of 
a thorough investigation of light propagation in the gravitational wave 
field generated by some isolated system. Our motivation is that electromagnetic 
waves are still the main carrier of astrophysically important information 
from very remote domains of our universe. Also, the operation of 
interferometric 
gravitational wave detectors and other techniques used for making 
experiments in gravitational physics (lunar laser ranging, very long 
baseline interferometry, pulsar timing, Doppler tracking, etc.) are 
fully based on the degree of our understanding of how light propagates in 
variable, 
time-dependent gravitational fields generated by various celestial bodies.
Although quite a lot of work has been done on this subject (see, for example, 
\cite{Br72}-\cite{KK92}) a real progress 
and much deeper insight into the nature of the problem has been achieved 
only recently \cite{KSG}-\cite{K97}. The main advantage of the integration 
technique which has been developed for finding the light-ray trajectory 
perturbed by the 
gravitational field is its account for the important physical property of 
gravitational radiation, namely, its retardation character. Previous 
authors, apart from Damour and Esposito-Far\`ese \cite{DE98}, accounted for 
the retardation of the gravitational field  
only in form of plane gravitational waves. Hence, effects produced in the 
near and induction zones of isolated astronomical sources emitting  
waves could not be treated in full detail. As a particular example of 
importance of such effects we note the problem of detection of 
gravitational waves created by g-modes of the Sun. The space 
interferometer LISA will be able to detect those waves. However, the 
problem is that LISA will fly in the induction zone of the emission process 
of these 
gravitational waves and, hence, a much more complete theoretical analysis of 
the working of the detector is needed. The 
approximation 
of a plane gravitational wave for the description of the detection procedure is 
definitely not sufficient. The other example could be effects caused by the
time-dependent gravitational field of the ensemble of binary stars in our 
galaxy. Timing of high-stable millisecond pulsars might be a tool for the
detection of stochastic effects produced by that field \cite{Kop99a}. 

\section{Wave generation from isolated sources}

\subsection{Einstein field equations}

The gravitational field is described in general relativity solely by the metric 
tensor $g_{\mu\nu}$ (and its inverse $g^{\mu\nu}$). It is generated by the stress-energy tensor of the matter fields $T^{\mu\nu}$ via the second-order differential equations 
\begin{equation}
\label{0}
R^{\mu\nu} -{1\over 2} g^{\mu\nu} R = {8\pi G\over c^4} \; T^{\mu\nu}\;,
\end{equation}
where $R^{\mu\nu}$ and $R=g_{\rho\sigma}R^{\rho\sigma}$ denote, respectively, 
the Ricci tensor and the Ricci scalar. We assume that the matter 
tensor $T^{\mu\nu}$ corresponds to an isolated source, i.e. $T^{\mu\nu}$ 
has a spatially compact support with maximal radius $a$, and that the 
internal gravity of the source is weak in the sense that its mass $M$ 
satisfies $G M \ll a c^2$. Within these conditions it is appropriate to 
write the metric $g^{\mu\nu}$ in the form of a small deformation of the flat 
metric $\eta^{\mu\nu} = {\rm diag} (-1,1,1,1)$. 
We pose $h^{\mu\nu} = \sqrt{-g} g^{\mu\nu} - \eta^{\mu\nu}$ 
($g$ = determinant of $g_{\mu\nu}$) and assume that 
each component of $h^{\mu\nu}$ is numerically small: 
$|h^{\mu\nu}|\ll 1$. We lower and raise all 
indices of our metric perturbation $h^{\mu\nu}$ with the flat metric; for instance,
$h_{\mu\nu} = \eta_{\mu\rho}\eta_{\nu\sigma} h^{\rho\sigma}$ and
$h = \eta^{\rho\sigma} h_{\rho\sigma}$. Then the field equations (\ref{0}) can 
be re-written in terms of the metric perturbation $h^{\mu\nu}$ by separating 
out a second-order linear operator acting on $h^{\mu\nu}$, and the remaining 
part of the equations, which is at least quadratic in $h^{\mu\nu}$ and its 
first and second derivatives, we conventionally set 
to the right side of the equations together with the matter tensor. This yields 
\begin{equation} 
\label{4}
\square \; h^{\mu\nu} -\partial^\mu H^\nu - \partial^\nu H^\mu +
\eta^{\mu\nu} \partial_\rho H^\rho = {16\pi G\over c^4} \; 
\tau^{\mu\nu}\;,
\end{equation} 
where $\square = \square_\eta$ denotes the flat d'Alembertian operator and where $H^\mu \equiv
\partial_\nu h^{\mu\nu}$; on the right side of the equation we have put
\begin{equation} 
\label{5}
\tau^{\mu\nu} = (-g) \; T^{\mu\nu} + {c^4\over 16\pi G}\; 
\Lambda^{\mu\nu}\;,
\end{equation}
which represents the total stress-energy distribution of both the matter fields -- first term in (\ref{5}) -- and the gravitational field itself -- second term involving the non-linear gravitational source $\Lambda^{\mu\nu}=O(h^2)$ (note that $\tau^{\mu\nu}$ transforms as 
a Minkowskian tensor under Lorentz transformations). The divergence of the 
left side of (\ref{5}) is identically zero by virtue of the 
Bianchi identities, therefore the pseudo-tensor $\tau^{\mu\nu}$ is conserved in the ordinary sense,
\begin{equation} 
\label{6}
\partial_\nu \tau^{\mu\nu} = 0 \;,
\end{equation}
which is equivalent to the covariant conservation of the matter tensor, 
$\nabla_\nu T^{\mu\nu} =0$.
A gauge transformation
$h^{\mu\nu} \rightarrow h^{\mu\nu} + \partial^\mu \xi^\nu + 
\partial^\nu \xi^\mu
- \eta^{\mu\nu} \partial_\lambda \xi^\lambda $ does not affect the left side 
of (\ref{4}), and consequently by solving for a vector $\xi^\mu$ 
the wave equation $\square \xi^\mu = - H^\mu$ one
can arrange that $h^{\mu\nu}$ satisfies the {\it harmonic-gauge} condition
$\partial_\nu h^{\mu\nu} = 0$. In this gauge the field equations (\ref{4}) 
simply become
\begin{equation} 
\label{7}
\square \; h^{\mu\nu} = {16\pi G\over c^4} \; \tau^{\mu\nu} \;.
\end{equation}

We want now to formulate the condition that the source 
is really isolated, i.e. it does not receive any radiation 
from other sources located far away, at infinity. Recall that 
we can express any homogeneous regular solution 
of the wave equation $\square h_{\rm hom}=0$ at a given field 
point in terms of the values of $h_{\rm hom}$ at some source 
points forming a surrounding surface at retarded times. 
This is the Kirchhoff formula (see e.g. \cite{Fock}), which 
reads in the case where the surrounding surface is a sphere,
\begin{equation}
\label{8}
h_{\rm hom} ({\bf x}',t') = \int\int {d\Omega\over 4\pi} 
\left[ {\partial\over \partial \rho} (\rho h_{\rm hom}) + {\partial\over c\partial t} 
(\rho h_{\rm hom}) \right] ({\bf x},t),
\end{equation}
where $\rho = |{\bf x} - {\bf x}'|$ and $t =t' -\rho/c$. To formulate the
no-incoming radiation condition we say that there should be no such 
homogeneous regular solutions $h_{\rm hom}$ (since they correspond physically 
to waves propagating from sources at infinity). Taking the limit 
$r\to +\infty$ with $t+r/c$= const. in Kirchhoff's formula, we then arrive at the physical conditions that
\begin{equation}
\label{9}
\lim_{\stackrel{r\to +\infty}{t+{r\over c}={\rm const.}}} \left[{\partial\over 
\partial r} (rh^{\mu\nu})+
{\partial\over c\partial t} (rh^{\mu\nu})\right] ({\bf x},t) = 0\;,
\end{equation}
and that $r \partial_\lambda h^{\mu\nu}$ should be bounded in this limit.
The no-incoming radiation condition (\ref{9}) is thus imposed at (Minkowskian) past null infinity ${\cal J}^-$ in a conformally rescaled space-time diagram.

\subsection{Multipole expansion in linearized gravity}\index{multipole expansion}

For the rest of this Section (and also in Section 5) we shall 
restrict ourselves to the case of linearized gravity, defined 
in particular by the neglect of the non-linear gravitational 
source term $\Lambda^{\mu\nu}$, for which the field equations 
in harmonic gauge $\partial_\nu h^{\mu\nu}=0$ read
\begin{equation}
\label{a1}
\square \; h^{\mu\nu} = {16\pi G\over c^4} \; T^{\mu\nu}\;.
\end{equation}
Within the linearized approximation the matter 
stress-energy tensor is divergenceless: $\partial_\nu T^{\mu\nu} =0$. 
Therefore the linearized approximation is inconsistent as 
regards the motion of the matter source, which in this 
approximation stays unaffected by the gravitational field. 
However this approximation is quite adequate for describing 
the generation of waves by a given source (for instance 
acted on by non-gravitational forces). From the no-incoming 
radiation condition (\ref{9}), 
we find that the unique solution of (\ref{a1}) is the retarded one:
\begin{equation}
\label{10}
h^{\mu\nu} ({\bf x},t) = -{4G\over c^4} \int 
{d^3{\bf x}'\over |{\bf x} -{\bf x}'|} \; T^{\mu\nu} 
({\bf x}',t -{1\over c} |{\bf x} -{\bf x}'|) \;.
\end{equation}
Since we are being interested in the wave-generation 
problem, we choose the field point outside the source, 
that is $r=|{\bf x}|>a$ (with the origin of the 
spatial coordinates at the center of the ball with 
radius $a$, so $a>|{\bf x}'|$), and we decompose 
in that region $h^{\mu\nu}$ into ``multipole moments''. 
The straightforward way to do this is to employ the 
standard Taylor formula for the formal limit ${\bf x}'\to 0$,
\begin{equation}
\label{11}
{T({\bf x}',t-|{\bf x}-{\bf x}'|/c)\over |{\bf x}-{\bf x}'|} =
\sum^{+\infty}_{l=0} {(-)^l\over l !} x'_L \partial_L
\left[ {T({\bf x}',t-r/c) \over r} \right]\;.
\end{equation}
Notice the short-hand notation $L=i_1i_2\cdots i_l$ for 
a multi-index with $l$ indices, as well as 
$x'_L ={x'}^L={x'}^{i_1} {x'}^{i_2}\cdots {x'}^{i_l}$,
$\partial_L =\partial_{i_1}\partial_{i_2}\cdots\partial_{i_l}$ where $\partial_i=\partial/\partial x^i$. From this Taylor 
expansion we immediately arrive at the following 
expression for the multipole decomposition of the metric perturbation,
\begin{equation}
\label{12}
{\cal M} (h^{\mu\nu}) ({\bf x},t) = -{4G\over c^4} \sum^{+\infty}_{l=0}
{(-)^l\over l !} \partial_L \left[ {1\over r} 
{\cal H}^{\mu\nu}_L (t-{r\over c})\right]\;, 
\end{equation}
where the ``multipole moments'' depend on the retarded 
time $u \equiv t-r/c$ and are given by 
\begin{equation}  
\label{13}
{\cal H}^{\mu\nu}_L (u) = \int
d^3 {\bf x}' x'_L T^{\mu\nu} ({\bf x}',u)\;,
\end{equation}
In (\ref{12}) we employ the notation ${\cal M}$ to distinguish the
multipole expansion ${\cal M} (h)$ from $h$ itself. Of course, we have 
numerically ${\cal M} (h)=h$ outside the source, however inside the 
source ${\cal M} (h)$ and $h$ will differ from each other; indeed $h$ is a 
smooth solution of the equations (\ref{10}) while ${\cal M} (h)$ satisfies 
$\square {\cal M} (h)=0$ and becomes infinite when $r\to 0$ [as it is clear 
from (\ref{12})]. In Section 4, dealing with the general case of the 
non-linear theory, we shall prefer to use the multipole expansion in terms 
of symmetric and trace-free (STF) multipole moments. 
In the present case it is simpler to use the non-STF 
moments ${\cal H}^{\mu\nu}_L$. A systematic investigation 
of the STF multipole expansion in linearized gravity can be found in \cite{DI91b}.

Applying the (linearized) conservation law $\partial_\nu T^{\mu\nu} =0$ 
we easily find certain physical evolution equations (and conservation laws) 
to be satisfied by the multipole moments (\ref{13}). Making use of the 
Gauss theorem to discard spatial divergences of compact-support terms we 
successively obtain (with $(n)$ referring to $n$ time-derivatives)
\begin{eqnarray}
\label{a3}
{1\over c} \stackrel{\!\!\!\!\!(1)}{{\cal H}^{\mu 0}_L} 
&=& l {\cal H}^{\mu (i_l}_{L-1)}\;,\\
\label{a4}
{1\over c} \stackrel{\!\!\!\!\!(2)}{{\cal H}^{00}_L} 
&=& l (l-1) {\cal H}^{(i_l 
i_{l-1}}_{L-2)}\;,
\end{eqnarray}
where the round brackets around spatial indices denote the 
symmetrization (and where $L-1=i_1\cdots i_{l-1}$; $L-2=i_1\cdots 
i_{l-2}$). As a consequence of (\ref{a3}) we see that the 
anti-symmetric part of ${\cal H}^{i0}_j$ in the indices $ij$ is constant. A more general consequence is 
\begin{equation}
\label{14}
{1\over c} \epsilon_{ijk} \stackrel{\!\!\!\!\!\!\!\!\!\!(1)}{{\cal H}^{j0}_{kL-1}} = (l-1) 
\epsilon_{ijk} {\cal H}^{j(i_{l-1}}_{L-2)k}\;.
\end{equation}
In the case of the lowest-order ($l=0$ and $l=1$) multipole moments the right sides 
vanish, and therefore these equations 
represent the conservation laws for the corresponding moments. Over all we find 
ten conservation laws, one for the mass-type monopole or total mass $M$, three for the 
mass-type dipole or center of mass position $X_i$ (times $M$), three for the 
time derivative of the mass dipole or linear momentum $P_i$, and three for the 
current-type dipole or total angular momentum $S_i$. Specifically, we define
\begin{alignat}{2}
\label{a5}
M & \equiv & {1\over c^2} {\cal H}^{00} & = \int d^3 {\bf x}\; {T^{00}\over c^2}\;,\\
P_i & \equiv & {1\over c} {\cal H}^{0i} & = \int d^3 {\bf x}\; {T^{0i}\over c}\;,\\
\label{aaa}
S_i & \equiv & \; {1\over c} \; \epsilon_{ijk}{\cal H}^{0k}_j & = 
\epsilon_{ijk} \int d^3 {\bf x}\; x_j {T^{0k}\over c}\;, \\
MX_i & \equiv & {1\over c^2} {\cal H}^{00}_{i}&  = \int d^3 {\bf x}\; x_i 
{T^{00}\over c^2}\;.
\end{alignat}
Then, from (\ref{a3})-(\ref{14}), we have 
\begin{equation}
\label{15}
{\dot M}=0,\qquad {\dot P_i}=0,\qquad 
{\dot S_i}=0,\qquad\hbox{and}\qquad{\dot X_i}={P_i\over M}\;.
\end{equation}

\section{The quadrupole moment formalism}\index{quadropole moment}

\subsection{Multipole expansion in the far region}

We analyze the gravitational field in the far-zone \index{far--zone}
of the source, in which we perform the expansion of the multipolar expansion 
${\cal M}(h^{\mu\nu}$) when $r\to +\infty$ with $t-r/c=$const 
(Minkowskian future null infinity 
${\cal J}^+$). To leading order $1/r$ the formula (\ref{12}) yields
\begin{equation}
\label{18}
{\cal M}(h^{\mu\nu}) = -{4G\over c^4r} \sum^{+\infty}_{l=0} {n_L\over c^l l!} 
\stackrel{\!\!\!\!(l)}{{\cal H}^{\mu\nu}_L}(u)
+ O \left({1\over r^2} \right)\;,
\end{equation}
[where $(l)$ represents the $l$th time-derivative and where 
$n_L\equiv n^L=n^{i_1}n^{i_2}\cdots n^{i_l}$ with $n^{i}=x^i/r$]. 
Because of the powers of $1/c$ in front of each multipolar piece, 
it is clear that the far-zone expansion of ${\cal M}(h)$ 
is especially useful when the numerical value of each term 
of the formula (\ref{18}) really scales with the factor 
$1/c^l$ in front of it. This will be the case when the 
typical velocities of the particles composing 
the system are small with respect to 
the speed of light, ${v/c}\equiv \varepsilon \ll 1$, or, 
equivalently, when the maximal radius $a$ of the system 
is much smaller than the wavelength $\lambda$ of 
the emitted gravitational radiation ($\lambda = cP$ 
where $P$ is the typical period of the internal motion). 
In particular, this ``slow motion'' assumption is 
always realized in the case of a self-gravitating 
system with weak internal gravity, for which we have 
\begin{equation}
\label{18'}
\varepsilon\equiv {v\over c}\sim {a\over\lambda}\sim \sqrt{G M\over c^2 a}\ll 1\;.
\end{equation}

Thus, for slowly-moving systems we can retain only the first few terms in 
the multipolar-post-Newtonian expansion (\ref{18}). 
Let us restrict ourselves to the terms
\begin{eqnarray}
\label{a7}
h^{00} &=& -{4G\over c^4r} \biggl\{ {\cal H}^{00} + {n_a\over c}
\stackrel{\!\!\!(1)}{{\cal H}^{00}_a} 
+ {n_{ab}\over 2c^2} \stackrel{\!\!\!\!(2)}{{\cal H}^{00}_{ab}}\biggr\}\;, \\\label{a8}
h^{0i} &=& -{4G\over c^4r} \biggl\{ {\cal H}^{0i} + {n_a\over c}
\stackrel{\!\!\!(1)}{{\cal H}^{0i}_a} \biggr\}\;, \\\label{a9}
h^{ij} &=& -{4G\over c^4r} \left\{ {\cal H}^{ij}\right\}\;.
\end{eqnarray}
For easier notation we do not indicate the multipole expansion 
${\cal M}$, nor the neglected $O(1/r^2)$ in the distance 
to the source. Using the conservation laws (\ref{a3})-(\ref{a4}) 
we can easily re-express the latter expressions in terms 
of the total mass $M$, total linear-momentum $P_i$, and 
the Newtonian quadrupole moment
\begin{equation}
\label{19}
Q_{ij} \equiv {1\over c^2} {\cal H}^{00}_{ij} 
= \int d^3 {\bf x} {T^{00}\over c^2} x_i x_j 
\;.
\end{equation}
Since $M$ and $P_i$ are conserved they do not participate 
to the radiation field which is therefore dominantly 
quadrupolar. Restoring the neglected post-Newtonian 
error terms we obtain 
\begin{eqnarray}
\label{a10}
h^{00} &=& -{4G\over c^2r} \biggl\{ M+ {n_a\over c} P_a + {n_{ab}\over 2c^2}
\stackrel{(2)}{Q}_{ab}(u) + O \left(\varepsilon^3\right)\biggr\}\;,
\\\label{a11}
h^{0i} &=& -{4G\over c^3r} \biggl\{ P_i + {n_a\over 2c} \stackrel{(2)}{Q}_{ai}(u) 
+ O \left(\varepsilon^2\right)\biggr\}\;,\\\label{a12}
h^{ij} &=& -{4G\over c^4r} \biggl\{ {1\over 2} \stackrel{(2)}{Q}_{ij}(u) + O \left(\varepsilon\right)\biggr\}\;.
\end{eqnarray}
Note that the contribution of the angular momentum $S_i$ appears only at the 
sub-dominant order $O(1/r^2)$ [see for instance (\ref{b22}) below]. 
When acting on terms of order $1/r$ in the distance like in 
(\ref{a10})-(\ref{a12}) the derivative $\partial_\nu$ is proportional to the 
(Minkowskian) null vector $k_\nu = (-1,\bf n)$; namely 
$\partial_\nu = - k_\nu \partial_0 + O(1/r^2)$. Using this, the harmonic 
gauge condition $\partial_\nu h^{\mu\nu}=0$ reads
\begin{equation}
\label{21}
k_\nu \stackrel{\!\!\!\!\!(1)}{h^{\mu\nu}} = O\left( {1\over r^2} \right)
\end{equation}
which is checked directly to be satisfied by the expressions 
(\ref{a10})-(\ref{a12}).

\subsection{The far-field quadrupole formula}

Using the gauge freedom $h^{\mu\nu} \rightarrow h^{\mu\nu} + \partial^\mu \xi^\nu + 
\partial^\nu \xi^\mu- \eta^{\mu\nu} \partial_\lambda \xi^\lambda$, we apply a gauge transformation to what is called the transverse-traceless (TT) gauge. Namely we pose
\begin{eqnarray}
\label{21a}
\xi^0 &=& {G\over 2rc^3}\biggl[ -n_{ab} \stackrel{(2)}{Q}_{ab}-\stackrel{(2)}{Q}_{aa}\biggr]\;,\\\label{21b}
\xi^i &=& {G\over 2rc^3}\biggl[ n_{iab} \stackrel{(2)}{Q}_{ab}
+n_i \stackrel{(2)}{Q}_{aa}-4n_a \stackrel{(2)}{Q}_{ia}\biggr]\;.
\end{eqnarray}
The new metric in TT coordinates, say $h_{\mu\nu}^{\rm TT}$ 
(where we lower indices with the flat metric), is straightforwardly 
seen to involve in its $00$ and $0i$ components only 
the static contributions of the mass monopole and dipoles, 
namely
\begin{equation}
\label{21c}
h_{00}^{\rm TT} = -{4GM\over c^2r} \left( 1+ {n_a\over c} 
\dot X_a\right)\;,\qquad\quad
h_{0i}^{\rm TT} = -{4G P_i\over c^3r}\;.  
\end{equation}
In the TT gauge the only radiating components of the 
field are the spatial ones, $ij$, and we obtain
\begin{equation}
\label{26}
h_{ij}^{\rm TT} = -{2G\over c^4r} P_{ijab}({\bf n})\biggl\{\stackrel{\!\!\!\!(2)}{I_{ab}}(u)+O\left(\varepsilon \right)\biggr\}
+ O \left({1\over r^2}\right) \;.
\end{equation}
The latter equation is known as the far-field quadrupole equation.
The TT projection operator is defined by 
$P_{ijab} = P_{ia} P_{jb} - {1\over 2} P_{ij} P_{ab}$, where
$P_{ij}=\delta_{ij}-n_in_j$. This is a projector: namely 
$P_{ijkl} P_{klab}=P_{ijab}$, onto the plane orthogonal to ${\bf n}$: 
thus for instance, $n_i P_{ijab} = 0$; and it is trace free: 
$P_{ijab} \delta_{ab}=0$, so we have substituted in (\ref{26}) 
the trace free part of the quadrupole moment $Q_{ab}$, i.e.
\begin{equation}
\label{26p}
I_{ij}=Q_{ij}-{\delta_{ij}\over 3} \; Q_{aa}+O\left(\varepsilon^2 \right)\;.
\end{equation}
We have added a remainder $O(\varepsilon^2)$ to indicate 
the post-Newtonian corrections in the source moment $I_L$ 
computed in Section 4 [note that the remainder in (\ref{26}) 
is only $O(\varepsilon)$]. 

As we see all the physical properties of the gravitational 
wave are contained into the TT projection (\ref{26}). As a 
consequence, the effects of the wave on matter fields are 
transverse: the motion of matter induced by the wave takes 
place only in the plane orthogonal to the 
propagation of the wave. Furthermore, from the trace-less 
character of the wave we see there can be only two independent 
components or polarization states. We introduce two 
polarization vectors, $\bf p$ and $\bf q$, 
in the plane orthogonal to the direction of propagation 
$\bf n$, forming an orthonormal right-handed triad. In terms of these 
polarization vectors the projector 
onto the transverse plane reads $P_{ij} =p_ip_j +q_iq_j$. 
The two polarization 
states (customarily referred to as the ``plus'' and ``cross'' 
polarizations) are defined by
\begin{equation}
\label{27}
h_+ = {p_ip_j-q_iq_j\over 2}\; h_{ij}^{\rm TT} \;, \qquad\quad h_\times = 
{p_iq_j+p_jq_i\over 2}\; h_{ij}^{\rm TT} \;.
\end{equation}

Until \hfil very \hfil recently \hfil all \hfil expectations \hfil was \hfil that \hfil any \hfil astrophysical \hfil (slowly-\\ moving) source would emit gravitational radiation according to 
(at least dominantly) the quadrupole formula (\ref{26}), 
involving the 
{\it mass-type} quadrupole moment $I_{ij}$. For instance 
the waves from the binary pulsar obey this formula. However, 
it has been realized by Andersson \cite{Andersson} and 
Friedman and Morsink \cite{Friedman} that in the case of 
the secular instability of the $r$-modes (rotation, or 
Rossby modes) of isolated newly-born neutron stars, the 
gravitational radiation is dominated by the variation of 
the {\it current-type} quadrupole moment $J_{ij}$. Here 
we give, without proof, the formula analogous to (\ref{26}) 
but for the current quadrupole:
\begin{equation}
\label{271}
h^{\rm TT}_{ij|{\rm current}} = \frac{8G}{3 c^5 r} \; 
P_{ijab}({\bf n})~\epsilon_{acd}~\! n_c \stackrel{\!\!\!\!(2)}{J_{bd}}(u)\;,
\end{equation}
where the (trace-free) current quadrupole moment is given by
\begin{equation}
\label{272}
J_{ij}=\epsilon_{ab(i} \int d^3 {\bf x}~x_{j)} x_a {T^{0b}\over c}
+O\left(\varepsilon^2 \right)\;.
\end{equation}
It could be, rather ironically, that the first direct 
detection of gravitational waves would follow the formula 
(\ref{271}) rather than the classic formula (\ref{26}) 
appearing in all text-books such as \cite{MTW}.

\subsection{Energy balance equation and radiation reaction}\index{radiation reaction}

The stress-energy pseudo-tensor of all matter and 
gravitational fields (in harmonic 
coordinates) $\tau^{\mu\nu}$ is defined by (\ref{5}). 
Now, for gravitational waves propagating in vacuum at 
large distances from their sources (in regions where 
the waves are almost planar), it is appropriate to define
the stress-energy tensor of the waves as the gravitational 
source term [involving $\Lambda ^{\mu\nu} (h)$] in the 
definition of $\tau^{\mu\nu}$, in which $h_{\mu\nu}$ 
is replaced by the far-field metric (\ref{26}). Since 
the expression of the metric is valid up to fractional 
terms $O(1/r^2)$ in the distance, and since 
$\Lambda^{\mu\nu}$ is at least quadratic in $h$, 
the stress-energy tensor of gravitational waves 
will be valid up to $O(1/r^3)$. 
Thus, we define, in the far-zone,
\begin{equation}
\label{22}
T^{\mu\nu}_{\rm GW} = {c^4\over 16\pi G} \; \Lambda^{\mu\nu} + 
O \left( {1\over r^3} \right)\;.
\end{equation}
Now to quadratic order $\Lambda^{\mu\nu}$ is a complicated sum 
of terms like $h\partial\partial h + \partial h\partial h$.
But when using (\ref{21}) together with the fact that 
$k^2=0$, this sum simplifies drastically and we end up 
with [still neglecting $O(1/r^3)$]
\begin{equation}
\label{23}
T^{\mu\nu}_{\rm GW} = {c^2\over 32\pi G}\; k^\mu k^\nu 
\stackrel{\!\!\!\!(1)}{h^{\rm TT}_{ij}} 
\stackrel{\!\!\!\!(1)}{h^{\rm TT}_{ij}}~\!= {c^2\over 16\pi G}\; k^\mu k^\nu \biggl[(\stackrel{\!\!(1)}{h_+})^2 + (\stackrel{\!\!(1)}{h_\times})^2\biggr]\;. 
\end{equation}
The second form is obtained from the inverse 
of (\ref{27}): $h^{\rm TT}_{ij}=(p_ip_j-q_iq_j)h_++(p_iq_j+p_jq_i)h_\times$. 
The expression (\ref{23}) takes the classic form 
$\sigma~k^\mu k^\nu$ of the stress-energy 
tensor for a swarm of massless particles (gravitons) 
moving with the speed of light. Notice from (\ref{23}) 
that the energy density of waves is positive 
definite.
In the general case where we do not neglect the terms 
$O(1/r^3)$ the previous expressions of 
$T^{\mu\nu}_{\rm GW}$ are still valid, 
but provided that one performs a suitable average 
over several gravitational wavelengths (see \cite{MTW}). 
For quadrupole waves, substituting the quadrupole formula 
(\ref{26}), we get
\begin{equation}
\label{29}
T^{\mu\nu}_{\rm GW} = {G\over 8\pi r^2c^4} \; 
k^\mu k^\nu P_{ijkl} ({\bf n}) 
\stackrel{(3)}{I}_{ij} \stackrel{(3)}{I}_{kl} \;.
\end{equation}

We can integrate the conservation law $\partial_\nu \tau^{\mu\nu}=0$
over the usual three-dimensional space (volume element $d^3 x$), 
and use the Gauss theorem to obtain a flux of $T^{\mu\nu}_{\rm GW}$ 
through a surface at infinity (exterior surface element $dS_i$), so that
\begin{equation}
\label{30}
{d\over dt} \int d^3 {\bf x}~ \tau^{\mu 0}= -c \int d S_i~\! T^{\mu i}_{\rm GW}\;.
\end{equation}
We consider the $\mu=0$ component of this law, substitute for 
$T^{\mu\nu}_{\rm GW}$ the expression (\ref{29}) at the 
quadrupole approximation, perform the angular integration 
assuming for simplicity a coordinate sphere at infinity (i.e. 
$dS_i=r^2 n_i d\Omega$), and obtain the famous Einstein (mass-type) 
quadrupole formula
\begin{equation}
\label{31}
{dE\over dt} = - {G\over 5c^5} \biggl\{
\stackrel{(3)}{I}_{ij}\stackrel{(3)}{I}_{ij}
+O\left(\varepsilon^2\right)\biggr\} \;,
\end{equation}
where $E=\int d^3x~ \tau^{00}$ represents the energy 
(matter + gravitation) of the source, and where we 
re-installed the correct post-Newtonian remainder 
$O(\varepsilon^2)$. Without proof we give also the 
formula corresponding to the current-type quadrupole moment,

\begin{equation}
\label{31aa}
\left({dE\over dt}\right)_{|{\rm current}} = - {16 G\over 45 c^7} 
\stackrel{(3)}{J}_{ij}\stackrel{(3)}{J}_{ij} \;,
\end{equation}
where the current quadrupole moment is defined by (\ref{272}).

Interestingly, we can treat the decrease of the energy 
as the result of the back-action
of a radiative force (cf. \cite{LL75}). We operate by parts the 
time-derivatives in (\ref{31}) so as to obtain
\begin{equation}
\label{32a}
{d\over dt} \left( E + {\delta E_5\over c^5} \right) = 
- {G\over 5c^5} \stackrel{(1)}{I}_{ij} 
\stackrel{(5)}{I}_{ij} 
\end{equation}
where we put on the left side a term in the form of a total time-derivative, 
representing a correction of order $1/c^5$ to the energy $E$, given by 
\begin{equation}
\label{32b}
{\delta E_5\over c^5}= {G\over 5c^5}\left[ \stackrel{(3)}{I}_{ij} 
\stackrel{(2)}{I}_{ij} -\stackrel{(4)}{I}_{ij} 
\stackrel{(1)}{I}_{ij} \right] \;.
\end{equation}
Now, after a time much longer than the characteristic 
period of the source (for definiteness one can consider a quasi-periodic 
source or perform a suitable average; see e.g. \cite{BRu81}), the 
contribution due to the correcting term (\ref{32b}) will become negligible 
as compared to the right hand side of (\ref{32a}). Therefore, 
{\it in the long term}, we can ignore this term and finally, the equation 
(\ref{32a}) can be re-written equivalently in a form where the 
energy loss in the source is the result of the work of a 
radiation reaction force ${\bf F}_{\rm reac}$, namely
\begin{equation}
\label{32c}
{dE\over dt} = - \int d^3 x~ {\bf F}_{\rm reac}\cdot{\bf v}
\end{equation}
where 
\begin{equation}
\label{32d}
F_{\rm reac}^i(t,{\bf x}) = {2G\over 5c^5}~\rho~x_j 
\stackrel{(5)}{I}_{ij} (t)\;, \qquad\rho\equiv T^{00}/c^2\;.
\end{equation} 
The equation (\ref{32c})-(\ref{32d}) is called the 
radiation-reaction quadrupole formula; the specific 
expression (\ref{32d}) of the radiation reaction force 
is called after Burke and Thorne \cite{Bu71,BuTh70,MTW}. 
This force is to be interpreted as a small Newtonian-like 
force superposed to the usual Newtonian force at the 2.5PN 
order (or $\varepsilon^5$).  
Actually, the Burke-Thorne radiation reaction force is valid only in a 
special gauge. That is, only in a special gauge, differing for instance 
from the harmonic or ADM gauges, does the source equation of motion involve 
at 2.5PN order the correcting force (\ref{32c})-(\ref{32d}). (See \cite{S83} 
for a discussion of various expressions of the radiation-reaction force in 
different gauges.) Notice that the reaction force 
(\ref{32d}) contains time derivatives up to the fifth order 
inclusively. In practice, for implementation in numerical codes, high 
time-derivatives have the tendency of decreasing the precision of a 
numerical computation, and therefore it is advantageous to choose other 
expressions of the reaction force for implementation in numerical codes 
\cite{BDS90,Rezzolla}. On the other hand, 
the order of second, third, and higher time-derivatives can be 
reduced by making use of the Newtonian equations of motion of the matter source. 
Subsequent implication of such a form 
of the radiation reaction in binary systems leads, for example, to 
the theoretical prediction of the rate of orbital decay shown in (\ref{1}).

\section{Post-Newtonian gravitational radiation}\index{post--Newtonian approximation}

\subsection{The multipole moments in the post-Newtonian approximation}

In Section 2 we presented the formula for the multipole expansion of the 
field outside the source in linearized gravity. In the present section let us present, 
without proof, 
the corresponding formula in the full {\it non-linear} theory, i.e. when 
the Einstein field equations (\ref{7}) are solved taking into account the 
gravitational 
source term $\Lambda^{\mu\nu}$. The formula will be valid whenever the 
post-Newtonian expansion is valid, i.e. when (\ref{18'}) holds. 
Under this assumption the field in the near-zone of a slowly-moving 
source can be expanded in non-analytic (involving logarithms) series of $1/c$ \cite{BD86}. 
The general structure of the expansion is
\begin{equation} 
\label{33}
{\overline h}^{\mu\nu} (t,{\bf x},c) = \sum_{p,q} {(\ln c)^q\over c^p}
\; h^{\mu\nu}_{pq} (t,{\bf x}) \;,
\end{equation} 
where $h^{\mu\nu}_{pq}$ are the functional coefficients of the expansion 
($p,q=$ integers, including the zero).
The general multipole expansion of the metric field ${\cal M}(h)$ is found 
by requiring 
that when re-developed in the near-zone in the limit of the parameter 
$r/c\to 0$ (which is 
equivalent with the formal re-expansion in the limit $c\to\infty$) it {\it matches} 
with the previous post-Newtonian expansion (\ref{33}) in the sense of the 
mathematical techniques of matched asymptotic expansions, i.e.
\begin{equation}
\label{34}
\overline {{\cal M}(h)}={\cal M}(\overline h) \;.
\end{equation} 
It is worthwhile noting that the equality (\ref{34}) should be true in the 
sense of formal series, i.e. term by term in each coefficient after both 
sides of the equation
are re-arranged with respect to the same expansion parameter.

We find \cite{B95,B98mult} that the multipole expansion generalizing (\ref{12}) 
to the full theory is composed of two terms,
\begin{equation}
\label{35}
{\cal M}(h^{\mu\nu}) = \hbox{finite part}\, \square^{-1}_R [{\cal M}(\Lambda^{\mu\nu})] 
- {4G\over c^4} \sum^{+\infty}_{l=0}
{(-)^l\over l!} \partial_L \left\{ {1\over r} {\cal
H}^{\mu\nu}_L (t-r/c) \right\}\;,
\end{equation} 
where $\square^{-1}_R$ is the inverse flat space-time retarded operator.
Herein, the first term is a particular solution of the Einstein field 
equations outside the matter compact support, i.e. it satisfies 
$\square h_{\rm part}^{\mu\nu}=\Lambda^{\mu\nu}$, and
the second term is a retarded solution of the source-free (homogeneous) 
wave equation, i.e. $\square h_{\rm hom}^{\mu\nu}=0$. The ``multipole 
moments'' parametrizing this homogeneous solution are given explicitly
by an expression similar to (\ref{13}),
\begin{equation}
\label{35a}
 {\cal H}^{\mu\nu}_L (u) = \hbox{finite part} 
\int d^3 {\bf x}~ x_L \,
{\overline \tau}^{\mu\nu}({\bf x}, u)\ ,   
\end{equation} 
but involving in place of the matter stress-energy tensor 
$T^{\mu\nu}$ the {\it post-Newtonian} expansion $\overline 
\tau^{\mu\nu}$, in the sense of (\ref{33}), of the {\it total} 
(matter+gravitation) pseudo-tensor $\tau^{\mu\nu}$ defined by (\ref{5}). 
Both terms in (\ref{35}) involve an operation of taking the finite part. 
This finite part can be defined precisely by means of an analytic 
continuation (see \cite{B98mult} for details), but in fact it is basically 
equivalent to 
taking the finite part of a divergent integral in the sense of 
Hadamard \cite{Hadamard}. Notice in particular that the finite part in the 
expression of the multipole moments (\ref{35a}) deals with the behaviour of 
the integral {\it at infinity}: $r\to\infty$ (without the finite part the 
integral would be divergent because of the factor $x_L\sim r^l$ in the 
integrand and the fact that the pseudo-tensor ${\overline \tau}^{\mu\nu}$ is 
not of compact support). One can show that the multipole expansion 
(\ref{35})-(\ref{35a}) is equivalent with a different one proposed recently 
by Will and Wiseman \cite{WWi96}.

Generally, it is more useful (for applications) to express the multipole 
expansion not in terms of the moments (\ref{35a}), but in terms of symmetric 
trace-free (STF) moments. We denote the STF projection with a hat,
$\hat x_L \equiv {\rm STF} (x^L)$, so that, for instance,
$\hat x_{ij} =x_i x_j-{1\over 3}\delta_{ij} {\bf x}^2$. 
Then it can be shown that the STF multipole expansion equivalent to 
(\ref{35})-(\ref{35a}) reads,
\begin{equation}
\label{36a}
{\cal M}(h^{\mu\nu}) = \hbox{finite part}\, \square^{-1}_R [
{\cal M}(\Lambda^{\mu\nu})] - {4G\over c^4} \sum^{+\infty}_{l=0}
{(-)^l\over l!} \partial_L \left\{ {1\over r} {\cal
F}^{\mu\nu}_L (t-r/c) \right\}\;,
\end{equation} 
where the parametrizing multipole moments are a bit more complicated,
\begin{equation}
\label{36b}
{\cal F}^{\mu\nu}_L (u) = \hbox{finite part} \int d^3 {\bf x}~
{\hat x}_L \int^1_{-1} dz~ \delta_l(z) {\overline \tau^{\mu\nu}} 
({\bf x}, u+z|{\bf x}|/c)\;. 
\end{equation}
With respect to the non-tracefree expression (\ref{35a}) this involves 
an extra integration over the variable $z$, with weighting function 
\begin{equation}
\label{36c}
\delta_l (z) = {(2l+1)!!\over 2^{l+1}l !} (1-z^2)^l\;, \qquad
\int^1_{-1} dz \delta_l (z) =1\;, \qquad \lim_{l\to +\infty} 
\delta_l(z) =\delta(z)\;. 
\end{equation}

The results (\ref{36a})-(\ref{36c}) permit us to define a 
very convenient notion of 
the {\it source multipole moments}. Quite naturally, these 
are constructed from the ten components of 
${\cal F}^{\mu\nu}_L (u)$. First of all, we reduce the 
number of independent components to only six by using 
the four relations given by the harmonic gauge condition 
$\partial_\nu h^{\mu\nu} = 0$. Next we apply standard 
STF techniques (see \cite{BD89,DI91b,B98mult} 
for details), and, in this way, we are able to 
define {\it six} STF-irreducible multipole moments, 
denoted $I_L, J_L, W_L, X_L, Y_L, Z_L$, which are 
given by {\it explicit} integrals extending over 
the post-Newtonian-expanded pseudo-tensor 
${\overline \tau}^{\mu\nu}$ like in (\ref{36b}). 
All of the moments $I_L, J_L, \cdots, Z_L$ are 
referred to as the moments of the source; however 
notice that, among them, only the moments $I_L$ 
(mass-type moment) and $J_L$ (current-type) 
play a physical role at the {\it linearized} level. 
The other four moments $W_L, X_L, Y_L, Z_L$ simply 
parametrize a linear gauge transformation and can often
be omitted from the consideration. Only at the order 
2.5PN or $\varepsilon^5$ do they start playing a physical 
role. The complete formulas for the moments $I_L, J_L$ are \cite{B98mult} 
\begin{eqnarray}
\label{il}
 I_L(u)&=& \hbox{finite part} \text{$\int$} d^3{\bf x}~\int^1_{-1} dz\biggl\{ \delta_l\hat x_L\Sigma -{4(2l+1)\over
  c^2(l+1)(2l+3)} \delta_{l+1} \hat x_{iL} \partial_t\Sigma_i
  \nonumber\\
 &&\qquad +{2(2l+1)\over
  c^4(l+1)(l+2)(2l+5)} \delta_{l+2} \hat x_{ijL}
  \partial_t^2\Sigma_{ij} \biggr\} ({\bf x},u+z |{\bf x}|/c)\;, \quad \\
\label{jl} 
J_L(u)&=& \hbox{finite part} \text{$\int$}
   d^3{\bf x}~\int^1_{-1} dz~\varepsilon_{ab<i_l} \biggl\{ \delta_l\hat
  x_{L-1>a} \Sigma_b   \nonumber\\ 
&&\qquad  -{2l+1\over c^2(l+2)(2l+3)} \delta_{l+1} \hat x_{L-1>ac}    
  \partial_t\Sigma_{bc}
  \biggr\} ({\bf x},u+z |{\bf x}|/c)\;.
\end{eqnarray}
In these expressions, $<>$ refers to the STF projection, and we have posed
\begin{equation}
\label{36p}
 \Sigma \equiv {\overline\tau^{00} +\overline\tau^{ii}\over c^2}\quad \hbox{(where $\overline{\tau}^{ii} \equiv\delta_{ij}\overline\tau^{ij}$)} 
\;,\qquad
 \Sigma_i \equiv {\overline\tau^{0i}\over c}\;,\qquad
\Sigma_{ij} \equiv \overline{\tau}^{ij}\;.
 \end{equation}
The moments $I_L, J_L$ given by these formulas are valid 
formally up to any post-Newtonian order. They constitute the 
generalization in the non-linear theory of the Newtonian 
moments introduced earlier in (\ref{26p}) and (\ref{272}). 

In order to apply usefully these moments to a given problem, one must find the 
explicit expressions of the moments at a given post-Newtonian order by 
inserting into them the components of the pseudo-tensor 
${\overline \tau}^{\mu\nu}$ obtained from an explicit post-Newtonian 
algorithm. Without entering into details, we find for instance 
that at the 1PN order the mass-type source 
moment $I_L$ is given (rather remarquably) by a simple 
compact-support formula \cite{BD86,B95}, on which we can, 
therefore, remove the finite part prescription: 
\begin{equation}
\label{37}
I_L=\text{$\int$} d^3{\bf x}\biggl\{\hat x_L\sigma 
+{|{\bf x}|^2\hat x_L\over 2c^2(2l+3)}
\partial_t^2\sigma - {4(2l+1)\hat x_{iL}\over
  c^2(l+1)(2l+3)}\partial_t\sigma_i \biggr\}+O\left(\varepsilon^4\right)\;.
\end{equation}
We denote the compact-support parts of the source scalar 
and vector densities in (\ref{36p}) by  
\begin{equation}
\label{38}
\sigma \equiv {T^{00}+T^{ii}\over c^2}\;,\qquad\sigma_i 
\equiv {T^{0i}\over c}\;.
\end{equation}
See Blanchet and Sch\"afer \cite{BS89} for application of the 
formula (\ref{37}) to the computation of the relativistic correction in 
the ${\dot P}_b$ of a binary pulsar [given to lowest order by (\ref{1})]. 
On the other hand, Damour, Soffel, and Xu \cite{DSX91} (extending previous 
work of Brumberg and Kopeikin \cite{K88}, \cite{BK89}) used the formula in their study 
of the Solar-system dynamics at 1PN order. The property of being 
of compact support is a special feature of the 1PN mass-moment $I_L$. 
To higher-order (2PN and higher) the mass moment $I_L$ is intrinsically 
of non-compact support (see its expression in \cite{B95}); hence the finite part 
prescription in the definition of the moment $I_L$ plays a crucial role at 
2PN. Similarly, starting already at 1PN order, the current-moment $J_L$ is 
intrinsically of non-compact support \cite{DI91a,B95}.

In a linear theory, the source multipole moments coincide evidently with the 
radiative multipole moments, defined as the coefficients of the multipole 
expansion of the $1/r$ term in the distance to the source at retarded 
times $t-r/c=$const. (this is evident from Section 2). However, in a 
non-linear theory like general relativity, the source multipole moments 
interact with each other in the exterior field through the non-linearities. 
This is clear from the presence of the first term 
in (\ref{36a}), containing the gravitational source $\Lambda^{\mu\nu}$, and 
which does contribute to the $1/r$ part of the metric at infinity. Therefore 
the source multipole moments must be related to the radiative ones, the latter 
constituting in this approach the actual observables of the field at infinity. 

In the TT projection of the metric field one can define two sets 
of radiative moments $U_L$ (mass-type) and $V_L$ (current-type). The {\it definition} of these moments is that they parametrize the $1/r$-term of the $ij$ components of the metric in TT gauge. Thus, extending the formula (\ref{26}), the radiative moments  are given by the decomposition
\begin{equation}
 h^{\rm TT}_{ij}=-{4G\over c^2r} P_{ijab} 
 \sum_{l \geq 2} {1\over c^l l !} \biggl\{ n_{L-2} U_{abL-2}
 - {2l \over c(l+1)} n_{cL-2}
 \varepsilon_{cd(a} V_{b)dL-2}\biggr\} 
+O\left( {1\over r^2}\right).
\end{equation}
The radiative moments $U_L,V_L$ are related to the $l$-th 
time-derivatives of the corresponding source moments. Let 
us give, without proof, the result of the connection of the 
radiative moments to the source moments (\ref{il})-(\ref{jl}) 
to the order $\varepsilon^3$ (or 1.5PN) inclusively. To this 
order some non-linear ``monopole-radiative $l$-pole'' interactions 
appear, which correspond physically to the scattering of the $l$-pole 
wave on the static curvature induced by the total mass of the 
source (i.e. the mass monopole $M\equiv I$) -- an effect well 
known under the name of tail of gravitational waves. We find \cite{BD92,B95}
\begin{eqnarray}
\label{b10}
U_L(u)&=& \stackrel{(2)}{I}_L(u)+ {2GM\over c^3} \int^{+\infty}_0 d
\tau \stackrel{(4)}{I}_L (u-\tau) \left[ \ln \left( {\tau\over 2b} \right)
+ \kappa_l\right]+ O\left(\varepsilon^4 \right), \\ \label{b11}
V_L(u)&=& \stackrel{(2)}{J}_L(u)+ {2GM\over c^3} \int^{+\infty}_0 d
\tau \stackrel{(4)}{J}_L (u-\tau) \left[ \ln \left( {\tau\over 2b} \right)+ 
\pi_l\right]+ O\left(\varepsilon^4 \right). \qquad
\end{eqnarray}
The same expressions come out from the Will and Wiseman formalism \cite{WWi96}.
Here, $b$ is a normalization constant (essentially irrelevant since it corresponds to a choice of the origin of time in the far zone), and $\kappa_l$, $\pi_l$ are given by 
\begin{equation}
\label{b12}
\kappa_l ={2l^2 +5l+4\over l(l+1)(l+2)} +
\sum^{l-2}_{k=1} {1\over k} \;,\qquad\quad 
\pi_l = {l-1\over l(l+1)} +
\sum^{l-1}_{k=1} {1\over k} \;. 
\end{equation}
The first term in (\ref{b10}) relates essentially to the original 
quadrupole formula (see (\ref{26})). From (\ref{b10})-(\ref{b11}) one sees that the first non-linearity in the propagation of the waves is at 1.5PN order 
with respect to the quadrupole formula. Thus, with enough precision, one can replace in (\ref{b10}) the mass-type moment $I_L$ by its compact-support expression that we computed in (\ref{37}) [since the post-Newtonian remainder in (\ref{37}) is $O(\varepsilon^4)$]. 

\subsection{Post-Newtonian radiation reaction}

Emission of gravitational radiation affects 
the equations of motion of an 
isolated system dominantly at the 2.5PN order 
beyond the Newtonian acceleration. 
In a suitable gauge the radiation-reaction force 
density at the 2.5PN order is given by the quadrupole 
formula (\ref{32d}). In this Section we extend this 
quadrupole formula to include the relativistic corrections 
up to the relative 1.5PN order, which means the absolute 4PN order 
with respect to the Newtonian force. The method is to compute 
the radiation reaction by means of the matching [in the sense of (\ref{34})]
of the post-Newtonian field to the exterior multipolar field. Indeed, 
recall that the post-Newtonian 
field is valid only in the near zone, and, thus, only via a matching can 
it incorporate information from the correct boundary condition,   
{\it viz} the no-incoming 
radiation condition imposed at infinity by equation (\ref{9}),
which specifies the braking character of gravitational radiation reaction.

To the relative 1.5PN order, and in a suitable gauge, it 
can be shown that the reaction force derives from 
some ``electromagnetic-like'' scalar and vector 
reaction potentials $V^{\rm reac}$ and $V^{\rm reac}_i$.
Explicitly we have \cite{B97reac}
\begin{eqnarray}
\label{b14}
V^{\rm reac}({{\bf x}}, t) &=& -{G \over 5c^5} x_{ij}
\biggl\{ \stackrel{(5)}{I}_{ij} (t) +{4GM \over c^3} \int^{+\infty}_0
d\tau \ln \left(\tau\over 2b \right) \stackrel{(7)}{I}_{ij}(t-\tau)\biggr\}\nonumber\\
&+&{G \over c^7}
\left[{1\over 189}x_{ijk} \stackrel{(7)}{I}_{ijk}(t) - {1\over 70} {{\bf x}}^2 x_{ij}
\stackrel{(7)}{I}_{ij}(t) \right]
+ O\left(\varepsilon^9 \right)\;,\\\nonumber  \\  \label{b15}
V^{\rm reac}_i ({{\bf x}}, t) &=& {G\over c^5}\left[{1\over 21} \hat x_{ijk}
\stackrel{(6)}{I}_{jk}(t)
- {4\over 45} \epsilon_{ijk}~x_{jm} \stackrel{(5)}{J}_{km}(t) \right] + 
O\left(\varepsilon^7\right)\;.
\end{eqnarray}
The dominant term in the formula (\ref{b14})-(\ref{b15}) is 
the standard Burke-Thorne reactive {\it scalar} potential at 
2.5PN order [compare (\ref{b14}) with (\ref{32d})]. In this 
term, consistently with the approximation, one must insert 
the 1PN expression of the moment as given by (\ref{37}). The Burke-Thorne 
term is of ``odd''-parity-type as it corresponds to an odd 
power of $1/c$, and thus changes sign upon a time
reversal (or more precisely when we replace the retarded 
potentials by advanced ones). 
Similarly is the next term in (\ref{b14})-(\ref{b15}) is 3.5PN 
(i.e. $\varepsilon^7$), involving both the mass-quadrupole, 
mass-octupole and current-quadrupole moments (the term 
$\varepsilon^5$ in the vector potential $V^{\rm reac}_i$
corresponds really to $\varepsilon^7$ in the equations 
of motion). However, notice that the next term in the 
reaction scalar potential $V^{\rm reac}$, at 4PN or 
$\varepsilon^8$ order, belongs to the ``even''-parity-type. 
Nevertheless this term is really part of the radiation 
reaction for it is not invariant under a time reversal, 
as is involves an integration over the ``past'' history of 
the source, so that when changing the retarded potentials 
to advanced ones the integration range would change to the 
whole ``future''; hence, this term does not 
stay invariant. It represents the contribution of 
tails in the radiation-reaction force, and is nicely 
consistent, in the sense of energy conservation, with 
the tails in the far zone [equation (\ref{b10})]. 
For explicit computations of the back-reaction to 
3.5PN order in the case of point-mass binary systems 
see Iyer and Will \cite{IW93,IW95}, and Jaranowski 
and Sch\"afer \cite{JaraS97}.

Using the matching (\ref{34}) one finds that the near-zone 
post-Newtonian metric (to 1.5PN 
relative order in both the ``damping'' and ``conservative'' 
effects) is parametrized in this gauge by some generalized potentials
\begin{equation}
\label{41}
{\cal V}_\mu=\square^{-1}_{\rm sym}\left[-4\pi G \sigma_\mu \right]+
V_\mu^{\rm reac}\;.
\end{equation}
The first term represents, to this post-Newtonian order, the conservative 
part of the metric; it is of the normal ``even''-parity-type and is given by 
the usual symmetric integral (half-retarded plus half-advanced) of the 
source densities $\sigma_\mu=(\sigma,\sigma_i)$ given by (\ref{38}). The 
second term $V_\mu^{\rm reac}$ denotes the radiation-reaction potentials 
(\ref{b14})-(\ref{b15}). 
By inserting the metric parametrized by (\ref{41}) into the  
equations of motion of the source (i.e. $\partial_\nu\tau^{\mu\nu}
=0\Leftrightarrow \nabla_\nu T^{\mu\nu}=0$), and considering the 
integral of energy, we obtain the balance equation
\begin{equation}
\label{42}
{d E\over dt} = \int d^3 {{\bf x}} \left\{ -\sigma \partial_t
  V^{\rm reac} + {4\over c^2} \sigma_j \partial_t V^{\rm reac}_j \right\}
  + O \left( \varepsilon^9\right)\;.
\end{equation}
Here $E$ denotes the energy of the source at the 1PN 
(or even 1.5PN) order. Actually what we obtain is not 
$E$ but some $E+\delta E_5/c^5+ \delta E_7/c^7$ like in (\ref{32a}). 
Arguing as before we neglect these $\delta E_5$ and $\delta E_7$. 
Substituting now the expressions (\ref{b14})-(\ref{b15}) 
for the reactive potentials (and neglecting other 
$\delta E$'s) we get
\begin{eqnarray}
\label{42a}
{dE\over dt} = &-&{G\over 5c^5}\left\{\stackrel{(3)}{I}_{ij}+{2GM \over c^3}
\int^{+\infty}_0 d\tau \ln
\left(\tau \over 2b\right) \stackrel{(5)}{I}_{ij}(t-\tau)\right\}^2\nonumber\\\
&-&{G\over c^7} \left[{1\over 189} \stackrel{(4)}{I}_{ijk} \stackrel{(4)}{I}_{ijk}
+{16 \over 45} \stackrel{(3)}{J}_{ij} \stackrel{(3)}{J}_{ij} \right] 
+O\left(\varepsilon^9\right)\;,
\end{eqnarray}
The right-side is exactly in agreement with the computation of the total flux energy emitted in gravitational waves at infinity, which is computed making use of the stress-energy 
tensor of gravitational waves (\ref{23}). In particular we recover in the brackets of the first term of (\ref{42a}) the third time-derivative of the radiative moment $U_{ij}$ including its tail contribution. 
The difference with the standard derivation of the flux is that instead of computing a surface integral at infinity we have performed the computation completely within the source, using the local source equations of motion.

\section{Light propagation in gravitational fields of isolated sources}\index{light propagation}

\subsection{General solution of the light propagation equation}

We are going now to calculate in linearized approximation the propagation 
of a light ray in the gravitational field of an isolated source at rest 
showing a mass-monopole, a spin-dipole and a time-dependent 
mass-quadrupole moment.  
In linearized approximation, the propagation equation for a 
particle (massless or with mass) with space coordinates $x^i(t)$ reads
\begin{eqnarray}
\label{b20}
\ddot{x}^{i}(t)&=&\frac{1}{2}g_{00,i}-g_{0i,t}-\frac{1}{2}g_{00,t}\dot{x}^i-
g_{ik,t}\dot{x}^k-\left(g_{0i,k}-g_{0k,i}\right)\dot{x}^k-\\
\nonumber&&\mbox{} 
g_{00,k}\dot{x}^k\dot{x}^i-\left(g_{ik,j}-\frac{1}{2}g_{kj,i}\right)
\dot{x}^k\dot{x}^j+
\left(\frac{1}{2}g_{kj,t}-g_{0k,j}\right)\dot{x}^k\dot{x}^j\dot{x}^i\;,
\end{eqnarray}
where the metric coefficients $g_{\mu\nu} = \eta_{\mu\nu} + f_{\mu\nu}$, 
in linear approximation, are related with $h^{\mu\nu}$ from Sections 2--4 through 
$f_{\mu\nu} = - h_{\mu\nu} + \frac{1}{2}\eta_{\mu\nu} h$. 
The dots denote differentiation with respect to 
time $t$ and $c=1$ has been put for simplicity.
In the linear approximation scheme, the velocity $\dot{x}^i$ 
appearing on the right-hand-side of (\ref{b20}) can be treated as 
a constant vector. For massless particles (photons) it has unit 
length, i.e. $\dot{x}^i \dot{x}^i = 1$. In the following we use
$\dot{x}^i = k^i$ in the right hand side of (\ref{b20}).   

The unperturbed motion of photons reads
\begin{equation}
\label{43}
x^i(t) = x_0^i + k^i (t-t_0)\;,
\end{equation}
where $x_0^i$ denotes the position of the photon at time of emission $t_0$. 
For solving the light propagation equation (\ref{b20}) it is very convenient
to introduce the new time parameter $\tau$ defined by $\tau = t - t^*$,
where $t^*$ denotes the time of the closest approach of the photon to the 
source of the gravitational field. Then it holds
\begin{equation}
\label{44}
x^i(\tau) = \xi^i + k^i \tau\;,
\end{equation}
where $\xi^i$ is the vector pointing from the position of the source to the 
position of the photon at the closest approach. Its length is the impact 
parameter $|\xi|=d$. Obviously, $\xi^i$ and 
$k^i$ are orthogonal to each other in the euclidean sense, 
i.e. $\xi^i k^i = 0$. Therefore, the 
length of $x^i$, $r=|x|$, takes the simple form $r=\sqrt{\tau^2 + d^2}$.
Introducing the derivatives $\hat{\partial}_i = \hat{P}_{ij}\partial / \partial\xi^j$
and $\hat{\partial}_{\tau} = \partial / \partial\tau$, where $\hat{P}_{ij} = \delta_{ij}
- k_ik_j$ is the projection operator onto the plane orthogonal to $k^i$, 
allows the light-propagation equation to be written as 
\begin{eqnarray}
\label{c1}
\ddot{x}^{i}(\tau)&=&\frac{1}{2}k^{\alpha}k^{\beta}
{\hat{\partial}}_i f_{\alpha\beta}-{\hat{\partial}}_{\tau}\left(
k^{\alpha}f_{i\alpha} + \frac{1}{2}k^if_{00}-\frac{1}{2}k^ik^j k^p 
f_{jp}\right)\;,
\end{eqnarray}
where the four-dimensional vector $k^{\alpha}$ reads $k^{\alpha} = (1,k^i)$.

To get a complete overview of the influence of a gravitational wave, 
emitted from an isolated source, on the propagation of light rays, we use 
the representation of the metric coefficients (\ref{12}) which is valid all-over 
in the space outside a domain which includes the matter source. 
Splitting $f_{\mu\nu}$ into a canonical part $f^{\rm can}_{\mu\nu}$  
which contains trace-free tensors only, and a gauge part, i.e.
$f_{\mu\nu} = f^{\rm can}_{\mu\nu} + \partial_{\mu}w_{\nu} + 
\partial_{\nu}w_{\mu}$, we obtain 
in case of mass-monopole, spin-dipole, and mass-quadrupole source moments
(remember the source being at rest)
\vspace{0.3 cm}
\begin{eqnarray}
\label{b21}
f_{00}^{\rm can}&=&\frac{2M}{r}+\partial_{pq}
\left[\frac{{I}_{pq}(t-r)}{r}\right]\;,\\\nonumber\\\label{b22}
f_{0i}^{\rm can}&=&-\frac{2\varepsilon_{ipq}{S}_p n_q}{r^2}+
2\partial_j\left[\frac{\dot{I}_{ij}(t-r)}{r}\right]\;,\\\nonumber\\
\label{b23}
f_{ij}^{\rm can}&=&\delta_{ij}f_{00}^{\rm can}+\frac{2}{r}\ddot{I}_{ij}(t-r)\;.
\end{eqnarray}
Herein, for simplicity, we have put $G=1$, and $\partial_i=
\partial/\partial x^i$. The mass $M$, spin $S^i$, and the 
quadrupole moment $I_{ij}$ of the source of gravitational waves are 
given (in the Newtonian approximation) 
by the expressions (\ref{a5}), (\ref{aaa}), and (\ref{26p}). 
The explicit expressions for the gauge functions
$w^{\mu}$ relating $f^{\rm can}_{\mu\nu}$ with $f_{\mu\nu}$
are important for general discussion of light propagation in the field of 
gravitational waves emitted by the isolated source. However, for the sake of 
simplicity they will be omitted. Their precise form can be found in \cite{KSG}. 

The insertion of these expressions (\ref{b21})-(\ref{b23}) into the equation 
(\ref{c1}) results in the equation 
\begin{eqnarray}
\label{b24}
\ddot{x}^i(\tau) & = & \left[2{M}\left({\hat{\partial}}_{i}-
k_i\hat{\partial}_{\tau}\right)-
2{S}^p\left(\varepsilon_{ipq}{\hat{\partial}}_{q\tau}
-k_j\varepsilon_{jpq}{\hat{\partial}}_{iq}
\right)\right]\biggl\{\frac{1}{r}\biggr\}\\ 
& & + \mbox{}\left({\hat{\partial}}_{ipq}-k_i{\hat{\partial}}_{pq\tau}+
2k_p{\hat{\partial}}_{iq\tau}\right)
\biggl\{\frac{I_{pq}(t-r)}{r}\biggl\}-
2\hat{P}_{ij}{\hat{\partial}}_{q\tau}\biggl\{\frac{\dot{I}_{jq}(t-r)}{r}\biggl\}\nonumber \\
& & - {\hat{\partial}}_{\tau\tau}\left[w^i+\varphi^i -k^i\;\left(w^0 + \varphi^0\right)\right]\;, \nonumber
\end{eqnarray}
where the vector $\varphi^{\mu}$ denotes terms which are of gauge type. 
The precise form of $\varphi^{\mu}$ is not important here and can be 
found in \cite{KSG}.

The solution of equation (\ref{b24}), using the boundary conditions 
$\dot{x}^i(-\infty) = k^i$ and $x^i(\tau_0)=x_0^i$ -- emission point
in space of the light ray at time $\tau_0$, reads 
\begin{eqnarray}
\label{b26}
\dot{x}^i(\tau)&=&k^i+\dot{\Xi}^i(\tau)\;,\\
\label{b27}
x^i(\tau)&=&x^i_N(\tau)+\Xi^i(\tau)-\Xi^i(\tau_0)\;,
\end{eqnarray}
where $x^i_N(\tau)$ denotes the unperturbed trajectory (\ref{44}). 
The relativistic perturbation of the trajectory is given by
\begin{eqnarray}
\label{b28}
\dot{\Xi}^i(\tau) &=&
\left(2{M}{\hat{\partial}}_{i}+2{
S}^pk_j\varepsilon_{jpq}{\hat{\partial}}_{iq}\right)A(\tau,{\bf{\xi}})+
{\hat{\partial}}_{ipq}B_{pq}(\tau,{\bf{\xi}})- \\ 
& & \mbox{} \left(2{M}k_i + 2{S}^p\varepsilon_{ipq}{\hat{\partial}}_q\right)
\biggl\{\frac{1}{r}\biggr\}-
\left(k_i{\hat{\partial}}_{pq}-
2k_p{\hat{\partial}}_{iq}\right)
\biggl\{\frac{I_{pq}(t-r)}{r}\biggl\} \nonumber \\ 
\nonumber && \mbox{} -
2\hat{P}_{ij}{\hat{\partial}}_{q}\biggl\{\frac{\dot{I}_{jq}(t-r)}{r}\biggl\} - 
{\hat{\partial}}_{\tau}\left[w^i+\varphi^i
-k^i\;\left(w^0+\varphi^0\right)\right]\;, \nonumber
\end{eqnarray}

\begin{eqnarray}
\label{b29}
\Xi^i(\tau)&=&\left(2{M}{\hat{\partial}}_{i}+2{
S}^pk_j\varepsilon_{jpq}{\hat{\partial}}_{iq}\right)B(\tau,{\bf{\xi}})-
\left(2{M}k_i+2{S}^p\varepsilon_{ipq}{\hat{\partial}}_q
\right)A(\tau,{\bf{\xi}}) \\
&&\mbox{} + {\hat{\partial}}_{ipq}D_{pq}(\tau,{\bf{\xi}})-
\left(k_i{\hat{\partial}}_{pq}-2k_p{\hat{\partial}}_{iq}
\right)B_{pq}(\tau,{\bf{\xi}})
-2\hat{P}_{ij}{\hat{\partial}}_{q}C_{jq}(\tau,{\bf{\xi}})-
\nonumber\\\nonumber&&\mbox{}
-w^i(\tau,{\bf{\xi}})-\varphi^i(\tau,{\bf{\xi}})
+k^i\;\left[w^0(\tau,{\bf{\xi}})+\varphi^0(\tau,{\bf{\xi}})\right]\;,
\end{eqnarray}
whereby the scalar functions $A$ and $B$ and derivatives of the 
tensors $B_{ij}, C_{ij}, D_{ij}$ are known fully explicitly. They are given by
\begin{align}
\label{45}
A(\tau,{\bf{\xi}}) & \equiv {\int}\frac{d\tau}{r}
={\int}
\frac{d\tau}{\sqrt{d^2+\tau^2}}=-\ln\left(%
\sqrt{d^2+\tau^2}-\tau\right)\;,
\\
\label{46}
B(\tau,{\bf{\xi}}) & \equiv {\int}A(\tau,{\bf{\xi}})d\tau=
-\tau \ln\left(\sqrt{d^2+\tau^2}-\tau\right)-\sqrt{d^2+\tau^2}\;,
\\
\label{47a}
{\hat{\partial}}_k B_{ij}(\tau,{\bf{\xi}}) & = 
\left(y r\right)^{-1}{I}_{ij}(t-r) \xi^k\;,
\\
\label{47b}
{\hat{\partial}}_k C_{ij}(\tau,{\bf{\xi}}) & = \left(y r\right)^{-1}
\dot{I}_{ij}(t-r) \xi^k\;,
\\
\label{47c}
{\hat{\partial}}_{ijk}
D_{pq}(\tau,{\bf{\xi}}) & = \frac{1}{y}\left[\left(\!\hat{P}^{ij} \! +\frac{\xi^i\xi^j}{y
r}\right)\!{\hat{\partial}}_k
B_{pq}(\tau,{\bf{\xi}})+\hat{P}^{jk}{\hat{\partial}}_i
B_{pq}(\tau,{\bf{\xi}})+\xi^j {\hat{\partial}}_{ik}
B_{pq}(\tau,{\bf{\xi}})\right],
\end{align} 
where the variable $y=\tau-\sqrt{\tau^2 + d^2}$ is the retarded time argument 
for the photon which passes through the point of closest approach to the 
source of the gravitational radiation at time $t^*=0$. More details concerning 
the method of calculation of light ray trajectory in time dependent gravitational
fields can be found in \cite{KSG,KS}.

\subsection{Time delay and bending of light}\index{time delay}\index{light deflection}

The time delay results in the form
\begin{eqnarray}
\label{c2}
t-t_0&=&|{\bf x}-{\bf x}_0|-{\bf{k}}\cdot{\bf{\Xi}}(\tau)+
{\bf{k}}\cdot{\bf{\Xi}}(\tau_0)\;,
\end{eqnarray}
or
\begin{eqnarray}
\label{c3}
t-t_0&=&|{\bf x}-{\bf x}_0|+\Delta_M(t,t_0)+\Delta_S(t,t_0)+
\Delta_Q(t,t_0)\;,
\end{eqnarray}
where $|{\bf x}-{\bf x}_0|$ is the usual Euclidean distance 
between the points of emission, ${\bf x}_0$, and reception, 
${\bf x}$, of the photon, $\Delta_M$ is the classical
Shapiro delay produced by the (constant) spherically symmetric part of 
the gravitational field of the deflector (see, e.g. \cite{MTW}), 
$\Delta_S$ is the Lense-Thirring or Kerr delay due to
the (constant) spin of the localized source of gravitational waves \cite{K97},
and  $\Delta_Q$ describes an additional
delay caused by the time dependent quadrupole moment of the source \cite{KSG}. 
Specifically we obtain
\begin{eqnarray}
\label{c4}
\Delta_M&=&2{ M} \ln\left[\frac{r+\tau}{r_0+\tau_0}\right]\;,
\\\label{c5}
\Delta_S&=&-2\varepsilon_{ijk}k^j{S}^k {\hat{\partial}}_i 
\ln\left[\frac{r+\tau}{r_0+\tau_0}\right]\;,
\\\label{c6}
\Delta_Q&=&{\hat{\partial}}_{ij}
\left[B_{ij}(\tau,{\bf{\xi}})-B_{ij}(\tau_0,{\bf{\xi}})\right]+
\delta_Q(\tau,{\bf{\xi}})-\delta_Q(\tau_0,{\bf{\xi}})\;,
\end{eqnarray}
where
\begin{eqnarray}
\delta_Q(\tau,{\bf{\xi}}) = k^i\left(w^i+\varphi^i\right)-w^0-\varphi^0\;.
\end{eqnarray}

Let us now denote 
by $\alpha^i$ the dimensionless vector describing the total angle of 
deflection of the light ray measured at the point of observation
and calculated with respect to
vector $k^i$ given at  past null infinity. It is defined according
to the relationship  
\begin{eqnarray}
\label{d1}
\alpha^i(\tau,{\bf{\xi}})&=&k^i [{\bf k}\cdot 
\dot{\bf{\Xi}}(\tau,{\bf{\xi}})]-\dot{\Xi}^i(\tau,{\bf{\xi}})
=-\;\hat{P}_{ij}\;\dot{\Xi}^j(\tau,{\bf{\xi}})\;.
\end{eqnarray}

For observers being far away from the source of the gravitational 
wave the projection of the mass-quadrupole tensor of the 
source of gravitational radiation onto the plane
orthogonal to the propagation direction of the gravitational wave is the 
crucial object which enters into the observable effects. It reads
\begin{equation}
\label{47}
{I}_{ij}^{\rm TT}=P_{ijpq} I_{pq}={I}_{ij}+
\frac{1}{2}\left(\delta_{ij}+n_i n_j\right)n_p n_q\; {
I}_{pq}-\left(\delta_{ip}n_j n_q+\delta_{jp}n_i n_q\right)\;{
I}_{pq}\;, 
\end{equation}
where again we denote $n_i=x^i/r$.

In the case of small impact parameter $d$ ($d/r_0\ll 1, d/r\ll 1$) we 
respectively obtain for the time delay and the angle of deflection 
\begin{equation}
\label{48}
t-t_0 - |{\bf x}-{\bf x}_0| = -4\psi+2M\ln(4r r_0)\;,
\end{equation}
\begin{equation}
\label{49}
\alpha_i=4{\hat{\partial}}_i\psi\;,
\end{equation}
where $\psi$ is the gravitational lens potential having the form 
\begin{eqnarray}
\psi&=&\left[{M}+\varepsilon_{jpq} k^p{S}^q{\hat{\partial}}_j+
\frac{1}{2}\;{I}_{pq}^{\rm TT}(t^{\ast})\;{\hat{\partial}}_{pq}
\right]\ln d\;.
\end{eqnarray}
(Notice that in this gravitational lens approximation $n^i = k^i$ holds.)
Remarquably, the gravitational lens potential does depend on the gravitational
source mass-quadrupole tensor only through its value at the time of closest 
approach. Furthermore, the gravitational lens potential decays like $1/d^2$,
i.e. it is not being influenced by the wave part of the gravitational field
\cite{DE98,KSG}. 

A direct consequence of the time delay formula is the frequency shift formula
for a moving gravitational source
\begin{equation}
\label{50}
\frac{\delta \nu}{\nu} = 4 \frac{\partial \psi}{\partial t^*} 
+ v^i \alpha^i\;,
\end{equation}
where $v^i$ is the velocity of the observer. It is worthwhile noting that 
the expression (\ref{50}) holds for the source of electromagnetic waves 
being at past null infinity. An exhaustive treatment of the gravitational 
frequency shift for arbitrary locations of observer, source of light, and 
the source of gravitational waves is rather complicated and has been done 
only recently in \cite{KS}.

\section{Detection of gravitational waves}

In the asymptotic regime of a gravitational wave field, the time delay 
reads
\begin{equation}
\label{51}
\Delta_Q({\bf{k}};t,t_0) = \frac{k^ik^j}{1-\cos  \theta}
\left[\frac{\dot{{I}}^{\rm TT}_{ij}(t-r)}{r} - 
\frac{\dot{{I}}^{\rm TT}_{ij}(t_0-r_0)}{r_0}\right]\;,
\end{equation}
where $\theta$ is the angle between the receiver - (light) emitter and 
receiver - (gravitational wave) source direction ($\cos  \theta = - 
N^ik^i$) and where the assumption 
$|{\bf{x}}-{\bf{x}}_0| \ll r$ has been made.
Let us now apply the above formula to the time delay in a Michelson 
interferometer. Obviously, in this case, $r=r_0$ holds. For simplicity we 
assume that the interferometer device is oriented orthogonal to the propagation 
direction of the gravitational 
wave, i.e. $N^ik^i_1 = N^ik^i_2$, where $k^i_1$ and $k^i_2$ denote the 
directions of the two interferometer arms which are taken to be orthogonal
($k^i_1k^i_2=0$) and of equal length $L$. We also assume that the light, 
emitted from the beam-splitter, is reflected once at the end mirrors.  
Furthermore, the interferometer arms are to be oriented such that they coincide
with the main axes of the plus-polarization. Then the relative time delay of 
the reflected light beams reads 
\begin{equation}
\label{52}
\Delta_Q({\bf{k}}_1;t, t-2L) = \frac{k^i_1k^j_1}{r_0}
[\dot{{I}}^{\rm TT}_{ij}(t-r_0) - \dot{{I}}^{\rm TT}_{ij}(t-r_0-2L)]\;.
\end{equation}
The multiplication of the relative time delay by the angular frequency of the 
laser light, $\omega$, which is treated as  
constant in the approximation under consideration, 
results in the measurable phase shift at time $t$ of 
$\Delta \Phi (t) = 2 \omega \Delta_Q({\bf{k}}_1; t, t-2L)$.
If we assume for the plus-component of the gravitational wave the expression
$h_+ (t-r_0) = A_+ \cos  (\omega_gt)$, where $\omega_g$ is the
constant frequency of the wave and $A_+$ its constant amplitude, 
we obtain for the phase shift (cf. \cite{S94})
\begin{equation}
\label{53}
\Delta \Phi (t) = 2A_+ \frac{\omega}{\omega_g} \sin (\omega_gL)
\cos [\omega_g(t- L)]\;. 
\end{equation}
The maximal amplitude is achieved if the condition $\omega_gL=\pi/2$ holds. 
This yields
\begin{equation}
\label{54}
\Delta_{\rm max} \Phi (t) = 
2A_+ \frac{\omega}{\omega_g} \sin (\omega_gt)\;.
\end{equation}

At the photo-diode the following photo-current results
\cite{Shoemaker} 
\begin{equation}
\label{55}
I_{\rm ph}(t) = I_{\rm min} + \frac{I_{\rm max}-I_{\rm min}}{2}\left[1-\cos  \phi(t)\right]\;,
\end{equation}
where the phase $\phi(t)$ is composed out of the signal $\Delta \Phi$
plus a modulation term from Pockels cells, $\phi_m \sin (\omega_mt)$, 
i.e.   
\begin{equation}
\label{56}
\phi(t) = \Delta \Phi(t) + \phi_m \sin (\omega_mt)\;.
\end{equation}
The approximate decomposition of the expression (\ref{55}) into a {\it dc} 
(``direct current'' or non-alternate) part and a signal part reads 
$I_{\rm ph}(t) = I_{\rm dc} + I_{\omega_m}$, where
\begin{equation}
\label{57}
I_{\rm dc} = I_{\rm min} + 
\frac{I_{\rm eff}}{2}\left[1-J_0(\phi_m)\cos  \Delta\Phi(t)\right]\;,
\end{equation}
\begin{equation}
\label{58}
I_{\omega_m} = I_{\rm eff} J_1(\phi_m) \sin \Delta\Phi(t) 
\sin (\omega_mt)\;,
\end{equation}
with Bessel functions $J_0$ and $J_1$; $I_{\rm eff} = I_{\rm max}-I_{\rm min}$.
Because of the smallness of the gravitational phase shift we get
\begin{equation}
\label{59}
I_{\rm dc} = I_{\rm min} + \frac{I_{\rm eff}}{2} \left[1-J_0(\phi_m)\right]\;,
\end{equation}
\begin{equation}
\label{60}
I_{\omega_m} = I_{\rm eff} J_1(\phi_m) \Delta\Phi(t) \sin (\omega_mt)\;.
\end{equation}
Using the equation (\ref{54}), the latter equation can be written
\begin{equation}
\label{61}
I_{\omega_m} = I_{\rm eff} J_1(\phi_m) A_+
 \frac{\omega}{\omega_g}
\biggl\{\cos\left[(\omega_m-\omega_g)t\right] - \cos\left[
(\omega_m+\omega_g)t\right]\biggr\}\;.
\end{equation}
In this side-band form, the signal from the gravitational wave is 
being detected. For more details we refer to \cite{S94}.


\begin{thebibliography}{0.8}
\bibitem{B91} D.G. Blair (ed.): {\it The Detection of Gravitational Waves} 
(Cambridge University Press, Cambridge 1991).

\bibitem{BBM62} H. Bondi, M.G.J. van der Burg, and A.W.K. Metzner,
{\it Proc. R. Soc. London} {\bf A 269}, 21 (1962).
\bibitem{P65} R. Penrose: {\it Proc. R. Soc.} (London) {\bf A 284}, 159 (1965).
\bibitem{EIH} A. Einstein, L. Infeld, and B. Hoffmann: {\it Ann. Math.} {\bf
39}, 65 (1938).
\bibitem{CE70} S. Chandrasekhar and F.P. Esposito: {\it Astrophys. J.}
{\bf 160}, 153 (1970).
\bibitem{Bu71} W.L. Burke: {\it J. Math. Phys.} {\bf 12}, 401 (1971).
\bibitem{BuTh70} W.L. Burke and K.S. Thorne: in {\it Relativity},
M. Carmeli et al. (eds.), Plenum Press: New York (1970) p. 208.
\bibitem{MTW} C.W. Misner, K.S. Thorne, and J.A. Wheeler,
{\it Gravitation}, Freeman: San Francisco (1973).
\bibitem{Ehl80} J. Ehlers: {\it Ann. N.Y. Acad. Sci.} {\bf 336}, 279 (1980).
\bibitem{PapaL81} A. Papapetrou and B. Linet: {\it Gen. Relat. Grav.} 
{\bf 13}, 335 (1981).
\bibitem{DD81a} T. Damour and N. Deruelle: {\it Phys. Lett.} {\bf 87A}, 81
(1981).
\bibitem{D83a} T. Damour, in {\it Gravitational Radiation}, N. Deruelle
and T. Piran (eds.), North-Holland: Amsderdam (1983), p. 59.
\bibitem{S85} G. Sch\"afer: {\it Ann. Phys.} (N.Y.) {\bf 161}, 81 (1985).
\bibitem{Kop85} S.M. Kopejkin (Kopeikin): {\it Astron. Zh.} {\bf 62}, 889 (1985), (In Russian).
\bibitem{Bo59} W.B. Bonnor: {\it Philos. Trans. R. Soc. London} {\bf A 251},
233 (1959).
\bibitem{Th80} K.S. Thorne: {\it Rev. Mod. Phys.} {\bf 52}, 299 (1980).
\bibitem{BD86} L. Blanchet and T. Damour: {\it Philos. Trans. R. Soc.
London} {\bf A 320}, 379 (1986).
\bibitem{BD89} L. Blanchet and T. Damour: {\it Ann. Inst. H. Poincar\'e}
(Phys. Th\'eorique) {\bf 50}, 377 (1989).
\bibitem{DI91a} T. Damour and B.R. Iyer: {\it Ann. Inst. H. Poincar\'e}
(Phys.  Th\'eorique) {\bf 54}, 115 (1991).
\bibitem{BD92} L. Blanchet and T. Damour: {\it Phys. Rev.} {\bf D 46},
4304 (1992).
\bibitem{B95} L. Blanchet: {\it Phys. Rev.} {\bf D 51}, 2559 (1995).
\bibitem{WWi96} C.M. Will and A.G. Wiseman: {\it Phys. Rev.} {\bf D 54},
4813 (1996).
\bibitem{PeM63} P.C. Peters and J. Mathews: {\it Phys. Rev.} {\bf 131},
435 (1963).
\bibitem{EH75} L.W. Esposito and E.R. Harrison: {\it Astrophys. J.}
{\bf 196}, L1 (1975).
\bibitem{Wag75} R.V. Wagoner: {\it Astrophys. J.} {\bf 196}, L63 (1975).
\bibitem{D83b} T. Damour: {\it Phys. Rev. Lett.} {\bf 51}, 1019 (1983).
\bibitem{TFMc79} J.H. Taylor, L.A. Fowler, and P.M. Mc Culloch: {\it 
Nature} {\bf 277}, 437 (1979).
\bibitem{T93} J.H. Taylor: {\it Class. Quantum Grav.} {\bf 10}, S167 (1993).
\bibitem{Br72}V.A. Brumberg,  
{\it Relativistic Celestial Mechanics}, Nauka: Moscow 
(1972). In Russian.
\bibitem{Brans}C.H. Brans: {\it Astrophys. J.} {\bf 197}, 1 (1975).  
\bibitem{MG}B. Mashhoon and L.P. Grishchuk: {\it Astrophys. J.} {\bf 236}, 990 (1980).  
\bibitem{Pol}A.G. Polnarev: {\it Sov. Astronomy} {\bf 29}, 607 (1985).
\bibitem{Hel}R.W. Hellings: {\it Astron. J.} {\bf 91}, 650; Erratum {\bf 92}, 
1446 (1986).
\bibitem{Br91}V.A. Brumberg,  
{\it Essential Relativistic Celestial Mechanics}, Adam 
Hilger: Bristol (1991).
\bibitem{Klion}S.A. Klioner: {\it Astron. Zh.} {\bf 68}, 1046 (1991).
In Russian.
\bibitem{KK92}S.A. Klioner and S.M. Kopeikin: {\it Astron. J.} {\bf 104}, 897 
(1992).
\bibitem{KSG} S.M. Kopeikin, G. Sch\"afer, C.R. Gwinn, and T.M. Eubanks: {\it 
Phys. Rev.} {\bf D 59}, 084023 (1999). 
\bibitem{KS} S.M. Kopeikin and G. Sch\"afer: {\it Phys. Rev.} {\bf D 60}, 
124002 (1999).
\bibitem{K97} S.M. Kopeikin: {\it J. Math. Phys.} {\bf 38}, 2587 (1997).
\bibitem{DE98} T. Damour and G. Esposito--Far\`ese: {\it Phys. Rev.} {\bf D 58},
042001 (1998). 

\bibitem{Kop99a}S.M. Kopeikin, {\it Timing Effects of Gravitational 
Waves from Localized Sources}, a talk given at the XXXIVth Rencontres de 
Moriond on ``Gravitational Waves and Experimental Gravity'', 
Les Arcs, 23-30 January 1999; e-print gr-qc/9903070.

\bibitem{Fock} V.A. Fock: {\it Theory of Space, Time and
Gravitation} (Pergamon, London 1959).

\bibitem{DI91b}T. Damour and B.R. Iyer: {\it Phys. Rev.} {\bf D 43},
3259 (1991).

\bibitem{Andersson} N. Andersson: {\it Astrophys. J.} {\bf 502}, 708 (1998).

\bibitem{Friedman} J.F. Friedman and S.M. Morsink: {\it Astrophys. J.} {\bf 502}, 714 (1998).

\bibitem{LL75} L.D. Landau and E.M. Lifshitz, {\it The Classical Theory of 
Fields} (Pergamon, New York 1975).

\bibitem{BRu81} R. Breuer and E. Rudolph: {\it Gen. Relat. Grav.} {\bf 13},
777 (1981).

\bibitem{S83} G. Sch\"afer: {\it Lett. Nuovo Cimento} {\bf 36}, 105 (1983).

\bibitem{BDS90} L. Blanchet, T. Damour, and G. Sch\"afer: {\it Mon. Not. R. Astr. Soc.} {\bf 242}, 289 (1990).

\bibitem{Rezzolla} L. Rezzolla, M. Shibata, H. Asada, T.W. Baumgarte, and S.L. Shapiro: {\it Astrophys. J.} {\bf 525}, 935 (1999).

\bibitem{B98mult} L. Blanchet: {\it Class. Quantum Grav.} {\bf 15}, 1971 (1998).

\bibitem{Hadamard} J. Hadamard: {\it Le probl\`eme de Cauchy et les 
\'equations aux d\'eriv\'ees partielles hyperboliques} (Hermann, Paris 1932).

\bibitem{BS89} L. Blanchet and G. Sch\"afer: {\it Mon. Not. R. Astr. Soc.}
{\bf 239}, 845 (1989); Erratum {\bf 242}, 704 (1990).

\bibitem{DSX91} T. Damour, M. Soffel, and C. Xu: {\it Phys. Rev.} {\bf D 43},
3273 (1991).

\bibitem{K88} S.M. Kopeikin: {\it Cel. Mechanics} {\bf 44}, 87 (1988).

\bibitem{BK89}V.A. Brumberg and S.M. Kopeikin: {\it Nuovo Cimento} {\bf B 103}, 63 (1989).

\bibitem{B97reac} L. Blanchet: {\it Phys. Rev.} {\bf D 55}, 714 (1997).

\bibitem{IW93} B.R. Iyer and C.M. Will: {\it Phys. Rev. Lett.} {\bf 70},
113 (1993).

\bibitem{IW95} B.R. Iyer and C.M. Will: {\it Phys. Rev.} {\bf D 52}, 6882 (1995).

\bibitem{JaraS97} P. Jaranowski and G. Sch\"afer: {\it Phys. Rev.} {\bf D 55}, 4712 (1997).

\bibitem{S94} P.R. Saulson: {\it Fundamentals of Interferometric Gravitational 
Wave Detection} (World Scientific, Singapore 1994).

\bibitem{Shoemaker} D. Shoemaker, R. Schilling, L. Schnupp, W. Winkler, 
K. Maischberger, and A. R\"udiger: {\it Phys. Rev.} {\bf{D 38}}, 423 (1988).

\end{thebibliography}
\end{document}